\documentclass[final,5p,twocolumn,times,sort&compress]{elsarticle}

\usepackage{mathrsfs}
\usepackage[latin1]{inputenc}
\usepackage{array}
\usepackage{dsfont}
\usepackage{amsmath}\usepackage{amssymb,stmaryrd}
\usepackage{booktabs}
\usepackage{lipsum}
\usepackage{cancel}
\usepackage{xcolor}
\usepackage{hyperref}
\usepackage[dvips]{feynmp}
\usepackage{graphics}
\usepackage{upgreek}
\usepackage{units}

\renewcommand{\eqref}[1]{\mbox{Eq.~(\ref{#1})}}

\newcommand{\figref}[1]{\mbox{Fig.~\ref{#1}}}
\newcommand{\secref}[1]{\mbox{Sec.~\ref{#1}}}

\begin{document}

\begin{frontmatter}

\title{Unitarity in Maxwell-Carroll-Field-Jackiw electrodynamics}

\author[ufma]{Manoel M. Ferreira, Jr.}\ead{manojr.ufma@gmail.com}
\author[cbpf]{Jose A. Helay\"{e}l-Neto}\ead{helayel@cbpf.br}
\author[biobio]{Carlos M. Reyes}\ead{creyes@biobio.cl}
\author[ufma]{Marco Schreck}\ead{marco.schreck@ufma.br}
\author[ufma]{Pedro D.S. Silva}\ead{pdiego.10@hotmail.com}

\address[ufma]{Departamento de F\'{i}sica, Universidade Federal do Maranh\~{a}o \\
Campus Universit\'{a}rio do Bacanga, S\~{a}o Lu\'{i}s -- MA, 65085-580, Brazil}

\address[cbpf]{Centro Brasileiro de Pesquisas F\'{i}sica, Rio de Janeiro -- RJ, 22290-180, Brazil}

\address[biobio]{Departamento de Ciencias B\'{a}sicas, Universidad del B\'{i}o B\'{i}o, Chill\'{a}n, 3800708, Chile}

\begin{abstract}

In this work we focus on the Carroll-Field-Jackiw (CFJ) modified electrodynamics in combination with
a {\em CPT}-even Lorentz-violating contribution. We add a photon mass term to the Lagrange density and
study the question whether this contribution can render the theory unitary. The analysis is based on
the pole structure of the modified photon propagator as well as the validity of the optical theorem.
We find, indeed, that the massive CFJ-type modification is unitary at tree-level. This result provides
a further example for how a photon mass can mitigate malign behaviors.

\end{abstract}

\begin{keyword}
Lorentz violation \sep Standard-Model Extension \sep Modified photons
\PACS 11.30.Cp \sep 12.60.-i \sep 14.70.Bh
\end{keyword}

\end{frontmatter}

\section{Introduction}

Lorentz symmetry violation has been of permanent interest in the past two decades
whereby the major part of the investigations have been carried out within the
minimal Standard-Model Extension (SME) \cite{Colladay,Samuel}. The minimal SME
incorporates power-counting renormalizable contributions for Lorentz violation
in all particle sectors and has been subject to various studies.

The {\em CPT}-even photon sector of SME has been investigated thoroughly with
the main objective to obtain stringent bounds on its 19 coefficients \cite{KM1}.
The absence of vacuum birefringence has led to bounds at the level of
$10^{-32}$ to $10^{-37}$ \cite{KM3} for the 10 birefringent coefficients.
Investigating vacuum Cherenkov radiation \cite{Cherenkov2} for
ultra-high energy cosmic rays \cite{Klink2,Klink3} has provided a set of tight
constraints on the remaining coefficients \cite{Kostelecky:2008ts}.

Furthermore, there has been a vast interest in understanding
the properties of the {\em CPT}-odd photon sector of the SME that is represented
by the Carroll-Field-Jackiw (CFJ) electrodynamics~\cite{Jackiw}. This theory has
been extensively examined in the literature with respect to its consistency \cite{Higgs},
modifications that it induces in quantum electrodynamics (QED) \cite{Adam,Soldati},
its radiative generation \cite{Radio}, and many other aspects. As vacuum
birefringence has not been observed, the related coefficients are strongly bounded
at the level of $\unit[10^{-43}]{GeV}$ \cite{Kostelecky:2008ts}.

It is known that the timelike sector of CFJ electrodynamics is plagued by several
problems such as negative contributions to the energy density \cite{Colladay} and
dispersion relations that become imaginary in the infrared regime. In particular,
it was shown that violations of unitarity are present, at least for small momenta
\cite{Adam}. In the current work, our objective is to study unitarity of the timelike
sector of CFJ electrodynamics.

We would like to find out whether the inclusion of a photon mass can, indeed, solve
these issues. This idea is not unreasonable, as it is well-known that the introduction
of a mass for an otherwise massless particle helps to get rid of certain problems.
For example, a photon mass can act as a regulator for infrared divergences.
Furthermore, adding a mass to the graviton renders gravity renormalizable
(despite being plagued by the Boulware-Deser ghost \cite{Boulware:1973my} that is
removed by the construction of de Rham, Gabadadze, and Tolley \cite{deRham:2010kj}).
It was indicated in \cite{Colladay} and demonstrated in~\cite{Cov_quant}
that a photon mass is capable of mitigating the malign behavior in CFJ
electrodynamics.

The paper is organized as follows. In \secref{sec:theoretical-setting} we
introduce the theory to be considered and discuss some of its properties.
We determine the photon polarization vectors as well as the modified photon
propagator in \secref{sec:polarization-vectors}. Subsequently, we express
the propagator in terms of the polarization vectors, which is a procedure
that was introduced in \cite{Cov_quant}. The resulting object turns out to be
powerful to obtain quite general statements on perturbative unitarity in
\secref{sec:unitarity}. In \secref{sec:application-electroweak} we briefly argue
how the previous results can be incorporated into the electroweak sector of the
SME. Finally, we conclude on our findings in \secref{sec:conclusions}. Natural
units with $\hbar=c=1$ are used unless otherwise stated.

\section{Theoretical setting}
\label{sec:theoretical-setting}

We start from a Lagrange density that is a Lorentz-violating modification of
the St\"{u}ckelberg theory:
\begin{align}
\label{eq:lagrange-density-theory}
{\cal L}_{\gamma}&=-\frac14 F^{\mu \nu}F_{\mu \nu}-\frac14 (k_F)_{\kappa \lambda \mu \nu} F^{\kappa  \lambda}F^{\mu \nu}
+\frac 12 (k_{AF})_{\kappa} \epsilon^{\kappa \lambda \rho \sigma}A_{\lambda} F_{\rho \sigma} \notag \\
&\phantom{{}={}}\,+\frac 12 m_{\gamma}^2 A^2-\frac {1}{2\xi}
(\partial\cdot A)^2\,,
\end{align}
with the \textit{U}(1) gauge field $A_{\mu}$, the associated field strength tensor $F_{\mu\nu}=\partial_{\mu}A_{\nu}-\partial_{\nu}A_{\mu}$, a photon mass $m_{\gamma}$, and a real parameter $\xi$. Lorentz violation is encoded in the {\em CPT}-even and {\em CPT}-odd background fields $(k_F)_{\kappa\lambda\mu\nu}$ and $(k_{AF})_{\kappa}$, respectively. All fields are defined on Minkowski spacetime with metric signature $(+,-,-,-)$. Furthermore, $\varepsilon^{\mu\nu\varrho\sigma}$ is the Levi-Civita symbol in four spacetime dimensions where we use the convention $\epsilon^{0123}=1$. Restricting \eqref{eq:lagrange-density-theory} to the first and third term only, corresponds to the theory originally investigated by Carroll, Field, and Jackiw in \cite{Jackiw}.

For a timelike choice of $k_{AF}$, the latter theory is known to have stability problems, which explicitly show up, e.g., in the corresponding energy-momentum tensor \cite{Colladay} or the dispersion relations \cite{Adam}. In analyses carried out in the past, the introduction of a photon mass as a regulator \cite{Cov_quant} turned out to resolve these issues. It prevents the dispersion relation from becoming imaginary in the infrared region, i.e., for low momenta. Note that the current upper limit for a photon mass is $\unit[10^{-27}]{GeV}$ \cite{Tanabashi:2018}. Although this constraint on a violation of \textit{U}(1) gauge invariance is very strict, it still lies many orders of magnitude above the constraints on the coefficients of CFJ theory. Finally, as the propagator of Proca theory is known to have a singularity for $m_{\gamma}\mapsto 0$, we also include the last term in \eqref{eq:lagrange-density-theory}, which was introduced by St\"{u}ckelberg.

Now, we employ a specific parameterization of the background fields as follows:
\begin{subequations}
\begin{align}
\label{eq:nonbirefringent-ansatz}
(k_F)_{\mu\nu\varrho\sigma}&=\frac{1}{2}(\eta_{\mu\varrho}\tilde{k}_{\nu\sigma}-\eta_{\mu\sigma}\tilde{k}_{\nu\varrho}-\eta_{\nu\varrho}\tilde{k}_{\mu\sigma}+\eta_{\nu\sigma}\tilde{k}_{\mu\varrho})\,, \displaybreak[0]\\[2ex]
\tilde{k}_{\mu\nu}&=2\zeta_1\left(b_{\mu}b_{\nu}-\eta_{\mu\nu}\frac{b^2}{4}\right)\,, \displaybreak[0]\\[2ex]
(k_{AF})_{\kappa }&=\zeta_2 b_{\kappa }\,,
\end{align}
\end{subequations}
with a symmetric and traceless $(4\times 4)$ matrix $\tilde{k}_{\mu\nu}$ and a four-vector $b_{\mu}$ that gives rise to a preferred direction in spacetime. If Lorentz violation arises from a vector-valued background field, a reasonable assumption could be that the latter is responsible for both {\em CPT}-even and {\em CPT}-odd contributions. Furthermore, $\zeta_1$ and $\zeta_2$ are Lorentz-violating coefficients of mass dimension 0 and 1, respectively, that are introduced to control the strength of {\em CPT}-even and {\em CPT}-odd Lorentz violation independently from each other. The parameterization of $(k_F)_{\mu\nu\varrho\sigma}$ stated in \eqref{eq:nonbirefringent-ansatz} is sometimes called the nonbirefringent \textit{Ansatz} \cite{Altschul:2006zz}, as it contains the 9 coefficients of $(k_F)_{\mu\nu\varrho\sigma}$ that do not provide vacuum birefringence at leading order in Lorentz violation.

Let us rewrite the Lagrange density of \eqref{eq:lagrange-density-theory} in terms of the preferred four-vector $b_{\mu}$ introduced before:
\begin{align}
{\cal L}_{\gamma}&=-\frac14aF^{\mu \nu}F_{\mu \nu}-\zeta_1 b_{\kappa }  b_{\mu } F^{\kappa}_{\phantom{\kappa}\lambda}F^{\mu \lambda} \notag \\
&\phantom{{}={}}+ \frac 1 2 \zeta_2  b_{\kappa}  \epsilon^{\kappa \lambda \rho \sigma}A_{\lambda} F_{\rho \sigma}+\frac 12 m_{\gamma}^2 A^2-\frac {1}{2\xi}
(\partial\cdot A)^2\,,
\end{align}
with $a=1-\zeta_1 b^2$. Theories with a similar structure can be generated by radiative corrections and were, for example, studied in \cite{andrianov1}. Performing suitable integrations by parts, the latter Lagrange density is expressed as a tensor-valued operator $\hat{M}^{\nu\mu}$ sandwiched in between two gauge fields according to ${\cal L}_{\gamma}=(1/2)A_{\nu}\hat{M}^{\nu\mu}A_{\mu}$ with
\begin{align}
\label{eq:operator-field-equations}
\hat M^{\nu \mu}&=\left[a\partial^2 + m_{\gamma}^2+2\zeta_1 (b\cdot \partial)^2\right]\eta^{\mu\nu}
 -\left(a-\frac{1}{\xi}\right)\partial^{\mu}\partial^{\nu} \notag \\
&\phantom{{}={}}-2\zeta_1  (b \cdot \partial) (b^{\mu}   \partial^{\nu}   + b^{\nu}   \partial^{\mu}) \notag \\
&\phantom{{}={}}+2\zeta_1\partial^2 b^{\mu}b^{\nu}+2\zeta_2
  \epsilon^{\alpha \nu \rho \mu}b_{\alpha}\partial_{\rho}\,.
\end{align}
This form directly leads us to the equation of motion for the massive gauge field:
\begin{equation}
\label{eq:field-equation-gauge}
\hat M^{\nu\mu}A_{\mu}=0  \,.
\end{equation}

\section{Polarization vectors}
\label{sec:polarization-vectors}

Analyzing the properties of the polarization vectors for modified photons will allow us to find a
relation between the sum over polarization tensors and the propagator~\cite{Cov_quant}. In turn, this
relation will be useful to compute imaginary parts of forward scattering amplitudes and to reexpress these
in terms of amplitudes associated with cut Feynman diagrams. The latter is required by the optical
theorem to test perturbative unitarity. Note that studies of Lorentz-violating modifications based on
the optical theorem have already been performed in several papers for field operators of mass dimension
4 \cite{Opt_Theorem_Minimal} as well as higher-dimensional ones \cite{Opt_Theorem_Nonminimal}.

The operator of Eq.~(\ref{eq:operator-field-equations}) transformed to momentum space (with a global sign
dropped) is given by
\begin{align}
\label{eq:field-eqs-operator-momentum}
M^{\nu\mu}&=\Big[ap^2-m_{\gamma}^2+2\zeta_1(b\cdot p)^2\Big]\eta^{\mu\nu}
 -\left(a-\frac{1}{\xi} \right)p^{\mu}p^{\nu} \notag \\
&\phantom{{}={}}-2\zeta_1  (b \cdot p) \left(b^{\mu}p^{\nu}+ b^{\nu}   p^{\mu}   \right ) \notag \\
&\phantom{{}={}}+2\zeta_1  p^2 b^{\mu}  b^{\nu}     +2\mathrm{i}\zeta_2  \epsilon^{\alpha \nu \rho \mu} b_{\alpha}  p_{\rho}\,.
\end{align}
We consider the eigenvalue problem
\begin{eqnarray}
\label{eq:eigenvalue-problem-polarization}
M^{\nu\mu} v^{(\lambda)}_{\mu}(p)= \Lambda_{\lambda}(p) v^{(\lambda)\nu} (p) \,,
\end{eqnarray}
for a basis $\{v^{(\lambda)}\}$ of polarization vectors which diagonalize the equation of motion.
We use $\lambda=\{0,+,-,3\}$ as labels for these vectors. The eigenvalue $\Lambda_{\lambda}(p)=\Lambda_{\lambda}$
corresponds to the dispersion equation of the mode $\lambda$. To find a real basis, we choose the temporal polarization vector as
\begin{eqnarray}
v^{(0)}_{\mu}=\frac{p_{\mu}}{\sqrt{p^2}}\,,
\end{eqnarray}
and the longitudinal one as
\begin{eqnarray}
v^{(3)}_{\mu}=\frac{p^2b_{\mu}-(p\cdot b)p_{\mu}}{\sqrt{p^2D}}\,,
\end{eqnarray}
with $D=(b\cdot p)^2-p^2b^2$, which is the Gramian of the two four-vectors $p^{\mu}$ and $b^{\mu}$.
It is not difficult to check that $p\cdot v^{(3)}=0$. The longitudinal mode becomes physical for a
nonvanishing photon mass. Let us choose $p^2>0$ such that we do not have to consider absolute values
of $D$ inside square roots.
The previous vectors are normalized according to
\begin{equation}
v^{(0)}_{\mu} v^{(0)\mu}=1\,,\quad v^{(3)}_{\mu} v^{(3)\mu}=-1\,.
\end{equation}
Proceeding with the evaluation of \eqref{eq:eigenvalue-problem-polarization} for $\lambda=0,3$ we obtain
\begin{subequations}
\begin{align}
\label{eq:dispersion-equations-0-3}
\Lambda_{0}&=\frac{p^2}{\xi}-m_{\gamma}^2\,,\quad \Lambda_3=Q-2\zeta_1D\,, \displaybreak[0]\\[2ex]
Q&=ap^2-m_{\gamma}^2+2\zeta_1(b\cdot p)^2\,.
\end{align}
\end{subequations}
The theory is gauge-invariant for vanishing photon mass. In this case, $\xi$ can be interpreted as a
gauge fixing parameter. The dependence of $\Lambda_0$ on $\xi$ tells us that this associated degree
of freedom is nonphysical. We will come back to this point later.

Now, we find the remaining two polarization states, which we label as $\lambda=\pm$.
First, let us define the two real four-vectors
\begin{subequations}
\label{eq:real-vectors}
\begin{align}
v^{(1)\mu}&=\epsilon^{\mu\nu\rho\sigma}\frac{p_{\nu}n_{\rho}b_{\sigma}}{N_1}\,, \displaybreak[0]\\[2ex]
v^{(2)\mu}&=\epsilon^{\mu\nu\rho\sigma}\frac{p_{\nu}v^{(1)}_{\rho}b_{\sigma}}{N_2}    \,,
\end{align}
\end{subequations}
with an auxiliary four-vector $n^{\mu}$. We normalize these vectors
as
\begin{equation}
v^{(1,2)}_{\mu}v^{(1,2)\mu}=-1\,,
\end{equation}
which fixes the normalization constants:
\begin{subequations}
\begin{align}
\label{eq:normalization-constant-1}
|N_1|^2&=|-p^2((n\cdot b)^2-n^2b^2) -n^2(b\cdot p)^2-b^2(p\cdot n)^2 \notag \\
&\phantom{{}={}}+2(p\cdot n)(b\cdot n)(p\cdot b)|\,, \\[2ex]
N_2^2&=D\,.
\end{align}
\end{subequations}
Both vectors of Eqs.~(\ref{eq:real-vectors}) are orthogonal to $p^{\mu}$ and $b^{\mu}$. Besides, they
satisfy
\begin{equation}
\label{eq:equation-vectors-v12}
M^{\nu\mu}v^{(1,2)}_{\mu}=Q \,v^{(1,2)\nu}\pm 2\mathrm{i}\zeta_2\sqrt{D}\,v^{(2,1)\nu}\,.
\end{equation}
We now introduce the linear combinations
\begin{equation}
v^{(\pm)}_{\mu}=\frac{1}{\sqrt 2}(v^{(2)}_{\mu}\pm\mathrm{i}v^{(1)}_{\mu})\,,
\end{equation}
that obey the properties
\begin{subequations}
\begin{align}
v^{(+)} \cdot v^{(+)}=v^{(-)}\cdot v^{(-)}&=0\,, \displaybreak[0]\\[2ex]
v^{(+)} \cdot v^{(+)*}=v^{(-)}\cdot v^{(-)*}&=-1\,.
\end{align}
\end{subequations}
Constructing suitable linear combinations of Eqs.~(\ref{eq:equation-vectors-v12}) results~in
\begin{subequations}
\begin{align}
M^{\nu\mu}v^{(\pm)}_{\mu}&=\Lambda_{\pm}v^{(\pm)\nu}\,, \displaybreak[0]\\[2ex]
\label{eq:dispersion-equations-plus-minus}
\Lambda_{\pm}&=Q\pm 2 \zeta_2 \sqrt{D}\,.
\end{align}
\end{subequations}
Hence, the dispersion equation for the transverse modes reads
\begin{equation}
\Lambda_+\Lambda_-=Q^2-4\zeta_2^2D=0\,.
\end{equation}
In analyses performed in the past, the sum over two-tensors formed from the polarization
vectors and weighted by the dispersion equations turned out to be extremely valuable
\cite{Cov_quant}. In particular,
\begin{equation}
\label{eq:decomposition-propagator}
P_{\mu\nu}=-\sum_{\lambda,\lambda'=0,\pm,3}g_{\lambda\lambda'}\frac{v^{(\lambda)}_\mu v^{*(\lambda')}_\nu}{\Lambda_{\lambda}}\,,
\end{equation}
with the dispersion equations $\Lambda_{0,3}$ of \eqref{eq:dispersion-equations-0-3}
and $\Lambda_{\pm}$ of \eqref{eq:dispersion-equations-plus-minus}. Inserting the explicit
expressions for the basis $\{v^{(\lambda)}\}$ results in
\begin{align}
\label{eq:propagator}
P_{\mu\nu}&=-\frac{Q}{\Lambda_+\Lambda_-}\eta_{\mu\nu}-\frac{2\mathrm{i}\zeta_2}{\Lambda_+\Lambda_-}\epsilon_{\mu\alpha\nu\beta}p^{\alpha}b^{\beta} \notag \\
&\phantom{{}={}}+\left[\frac{(p\cdot b)^2}{(Q-2\zeta_1D)p^2D}-\frac{\xi}{p^2(p^2-\xi m_{\gamma}^2)}-\frac{Qb^2}{D\Lambda_+ \Lambda_-}\right]p_{\mu}p_{\nu} \notag \\
&\phantom{{}={}}+\frac{(p\cdot b)}{D}\left(\frac{Q}{\Lambda_+\Lambda_-}-\frac{1}{Q-2\zeta_1D}\right)(p_{\mu}b_{\nu}+p_{\nu}b_{\mu}) \notag \\
&\phantom{{}={}}+\frac{p^2}{D}\left(\frac{1}{Q-2\zeta_1D}-\frac{Q}{\Lambda_+\Lambda_-}\right)b_{\mu}b_{\nu}\,.
\end{align}
By computation, we showed that $P_{\mu\nu}$ is equal to the negative of the inverse of
the operator $M^{\mu\nu}$ in \eqref{eq:field-eqs-operator-momentum}:
$M^{\mu\nu}P_{\nu\varrho}=-\delta^{\mu}_{\phantom{\mu}\varrho}$. Therefore, we define
$\mathrm{i}P_{\mu\nu}$ as the modified photon propagator of the theory given by
\eqref{eq:lagrange-density-theory}. For $\zeta_1=\zeta_2=0$ and $m_{\gamma}\mapsto 0$
we observe that
\begin{equation}
\mathrm{i}P_{\mu\nu}=-\frac{\mathrm{i}}{p^2}\left[\eta_{\mu\nu}+(\xi-1)\frac{p_{\mu}p_{\nu}}{p^2}\right]\,,
\end{equation}
which is the standard result for the propagator in Fermi's theory, as expected.

\section{Perturbative unitarity at tree level}
\label{sec:unitarity}

Now were are interested in studying probability conservation for the theory given
by \eqref{eq:lagrange-density-theory}. It is known that the timelike sector of CFJ
theory has unitarity issues \cite{Adam} while it was also demonstrated
that a photon mass helps to perform a consistent quantization \cite{Cov_quant}.
Therefore, we would like to investigate the question whether the presence of a
photon mass can render the theory unitary. For brevity, we choose a purely timelike
background field: $(b^{\mu})=(1,0,0,0)$. Note that in this case, the theory
of \eqref{eq:lagrange-density-theory} is isotropic and we can identify
$\zeta_1$ with the isotropic {\em CPT}-even coefficient
$\tilde{\kappa}_{\mathrm{tr}}$ \cite{Kostelecky:2002hh}. For completeness, we
keep the {\em CPT}-even contributions, although they are not expected to cause unitarity
issues for small enough $\zeta_1$.

In the classical regime, a four-derivative applied to the field
equations (\ref{eq:field-equation-gauge}) provides
\begin{equation}
\left(\frac{1}{\xi}\square+m_{\gamma}^2\right)\partial\cdot A=0\,,
\end{equation}
where $\square=\partial^{\mu}\partial_{\mu}$ is the d'Alembertian.
Even if $A_{\mu}$ couples to a conserved current, $\partial\cdot A$ behaves as
a free field. It can be interpreted as a nonphysical scalar mode that exhibits
the dispersion relation
\begin{equation}
\label{eq:dispersion-relation-scalar-mode}
\omega_k^{(0)}=\sqrt{\vec{k}^{\,2}+\xi m_{\gamma}^2}\,.
\end{equation}
Therefore, in the classical approach, the St\"{u}ckelberg term proportional to
$1/\xi$ can actually be removed from the field equations. In this case, we
automatically get the subsidiary requirement $\partial\cdot A=0$ that corresponds
to the Lorenz gauge fixing condition.
\begin{figure}[t]
\centering
\includegraphics[width=0.3\textwidth]{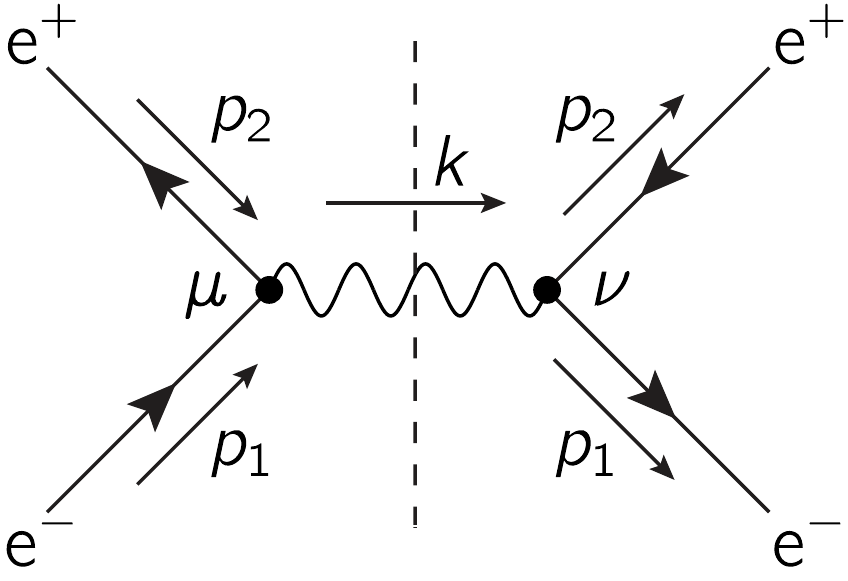}
\caption{Electron and positron forward scattering with the dashed line representing
a cut of the diagram.}
\label{fig:Fig1}
\end{figure}

Furthermore, the modified Gauss and Amp\`{e}re law for the
electric field $\vec{E}$ and magnetic field $\vec{B}$ are obtained directly
from \eqref{eq:field-equation-gauge} and read as follows:
\begin{subequations}
\begin{align}
(1+\zeta_1)\vec{\nabla}\cdot\vec{E}+m_{\gamma}^2\phi&=0\,, \displaybreak[0]\\[2ex]
(1-\zeta_1)\vec{\nabla}\times\vec{B}-2\zeta_2\vec{B}+m_{\gamma}^2\vec{A}&=(1+\zeta_1)\partial_t\vec{E}\,,
\end{align}
\end{subequations}
where $\phi$ is the scalar and $\vec{A}$ the vector potential, respectively.
By expressing the physical fields in terms of the potentials and using the Lorenz
gauge fixing condition, the modified Gauss law can be brought into the form
\begin{equation}
\left[(1+\zeta_1)(\partial_t^2-\triangle)+m_{\gamma}^2\right]\phi=0\,,
\end{equation}
with the Laplacian $\triangle=\vec{\nabla}^2$. This massive wave
equation leads to the dispersion relation
\begin{equation}
\label{eq:dispersion-relations-longitudinal}
\omega_k^{(3)}=\sqrt{\vec{k}^{\,2}+\frac{m_{\gamma}^2}{1+\zeta_1}}\,.
\end{equation}
The associated mode is interpreted as longitudinal. Furthermore, the
modified Amp\`{e}re law yields
\begin{align}
\vec{0}&=\left[(1+\zeta_1)\partial_t^2-(1-\zeta_1)\triangle+m_{\gamma}^2\right]\vec{A}-2\zeta_1\vec{\nabla}(\vec{\nabla}\cdot\vec{A}) \notag \\
&\phantom{{}={}}-2\zeta_2\vec{\nabla}\times\vec{A}\,.
\end{align}
The latter provides the modified transverse dispersion relations
\begin{equation}
\label{eq:dispersion-relations-transverse}
\omega_k^{(\pm)}=\sqrt{\frac{(1-\zeta_1)\vec{k}^{\,2}\mp 2\zeta_2|\vec{k}|+m_{\gamma}^2}{1+\zeta_1}}\,.
\end{equation}
Note that the dispersion relations found above correspond to the poles of
the propagator that we have calculated in \eqref{eq:propagator}. These results
will be valuable in the quantum treatment to be performed as follows.

First, we would like to construct the amplitude of a scattering process
involving two external, conserved four-currents $J_{\mu}$, $J_{\nu}^{*}$
without specifying them explicitly:
\begin{equation}
\mathcal{S}\equiv J^{\mu}(\mathrm{i}P_{\mu\nu})J^{*,\nu}\,.
\end{equation}
The latter object is sometimes called the saturated propagator in the literature
\cite{Veltman:1981,SaturatedPropagator} (cf.~\cite{Nakasone:2009bn} for an
application of this concept in the context of massive gravity in $(1+2)$ dimensions).
Inserting the decomposition (\ref{eq:decomposition-propagator}) of the propagator in terms of
polarization vectors, we obtain:
\begin{equation}
\mathcal{S}=-\mathrm{i}\sum_{\lambda=\pm,3} \frac{|J\cdot v^{(\lambda)}|^2}{\Lambda_{\lambda}}\,,
\end{equation}
where the mode labeled with $\lambda=0$ is eliminated because of current conservation: $p\cdot J=0$.
To guarantee the validity of unitarity, the imaginary part of the residues of $\mathcal{S}$
evaluated at the positive poles in $p_0$ should be nonnegative. The numerator of the latter expression
is manifestly nonnegative. Hence, the outcome only depends on the pole structure of $\mathcal{S}$.
For the case of a purely timelike $b_{\mu}$, which is the interesting one to study, we obtain
\begin{equation}
\label{eq:imaginary-part-residues}
\mathrm{Im}[\mathrm{Res}(\mathcal{S})|_{k_0=\omega_k^{(\lambda)}}]=\frac{1}{2(1+\zeta_1)}\frac{|J\cdot v^{(\lambda)}|^2}{\omega_k^{(\lambda)}}\,,
\end{equation}
for each one of the physical modes $\lambda=\pm,3$ with dispersion
relations~(\ref{eq:dispersion-relations-longitudinal}),
(\ref{eq:dispersion-relations-transverse}).
For perturbative $\zeta_1$, $\zeta_2$, and $m_{\gamma}>|\zeta_2|/\sqrt{1-\zeta_1}$ it
holds that $\omega_k^{(\lambda)}>0$ for $\lambda=\pm,3$. In this case, the right-hand
side of \eqref{eq:imaginary-part-residues} is nonnegative. This result is already an
indication for the validity of unitarity.
We see how a decomposition of the form of \eqref{eq:decomposition-propagator}
leads to a very elegant study of the saturated propagator without the need of
considering explicit configurations of the external currents such as done in the
past \cite{SaturatedPropagator}.

To test unitarity more rigorously, we couple \eqref{eq:lagrange-density-theory}
to standard Dirac fermions and consider the theory
\begin{subequations}
\begin{align}
\mathcal{L}&=\mathcal{L}_{\gamma}+\mathcal{L}_{\psi,\gamma}\,, \\[2ex]
\mathcal{L}_{\psi,\gamma}&=\overline{\psi}[\gamma^{\mu}(\mathrm{i}\partial_{\mu}-eA_{\mu})+m_{\psi}]\psi\,.
\end{align}
\end{subequations}
Here, $e$ is the elementary charge, $m_{\psi}$ the fermion mass, $\gamma^{\mu}$ are
the standard Dirac matrices satisfying the Clifford algebra $\{\gamma^{\mu},\gamma^{\nu}\}=2\eta^{\mu\nu}$,
$\psi$ is a Dirac spinor, and $\overline{\psi}=\psi^{\dagger}\gamma^0$ its Dirac conjugate.
We intend to check the validity of the optical theorem for the tree-level forward scattering
amplitude of the particular electron-positron scattering process in \figref{fig:Fig1}.
The amplitude ${\cal M}_F$ associated with the corresponding Feynman graph is
\begin{align}
\label{Amplitud}
\mathrm{i}{\mathcal M}_F&=\bar{v}^{(r)}(p_{2})(-\mathrm{i}e\gamma^{\mu})u^{(s)}(p_{1})(\mathrm{i}P^F_{\mu \nu }(k)) \notag \\
&\phantom{{}={}}\times\bar{u}^{(s)}(p_{1})(-\mathrm{i}e\gamma^{\nu})v^{(r)}(p_{2})\,,
\end{align}
with particle spinors $u^{(s)}$ and antiparticle spinors $v^{(s)}$ of spin projection $s$.
The momentum of the internal photon line is $k=p_1+p_2$. Furthermore,
$P^F_{\mu\nu}(k)$ is the Feynman propagator obtained from \eqref{eq:decomposition-propagator}
by employing the usual prescription $k^2\mapsto k^2+\mathrm{i}\epsilon$. Let us define the
four-currents
\begin{subequations}
\label{Currents}
\begin{align}
J^{\mu}&\equiv\bar{v}^{(r)}(p_{2})\gamma ^{\mu}u^{(s)}(p_{1})\,, \displaybreak[0]\\[2ex]
J^{\ast\mu}&\equiv\bar{u}^{(s)}(p_{1})\gamma ^{\mu}v^{(r)}(p_{2})\,.
\end{align}
\end{subequations}
Due to current conservation at the ingoing and outgoing vertices, $p\cdot J=p\cdot J^{*}=0$.
Introducing an integral and a $\delta$ function for momentum conservation, we can write the
forward scattering amplitude as
\begin{equation}
\label{eq:forward-scattering-amplitude-general}
{\mathcal M}_F=-e^2 J^{\mu }J^{\ast\nu}\int\frac{\mathrm{d}^4k}{(2\pi)^4}\,P^F_{\mu \nu}(k)(2\pi)^4\delta^{(4)}(k-p_1-p_2)\,.
\end{equation}
Now we insert \eqref{eq:decomposition-propagator} and decompose the denominator in terms of
the poles as follows:
\begin{align}
\label{eq:amplitude-propagator-inserted}
\mathcal{M}_F&=-e^2 J^{\mu}J^{\ast\nu}\int\frac{\mathrm{d}^4k}{(2\pi)^4}\,\sum_{\lambda,\lambda'=0,\pm,3}\frac{-g_{\lambda\lambda'}}{1+\zeta_1} \notag \\
&\phantom{{}={}}\times\frac{v_{\mu}^{(\lambda)}v_{\nu}^{*(\lambda')}}{(k_0-\omega_k^{(\lambda)}+\mathrm{i}\epsilon)(k_0+\omega_k^{(\lambda)}-\mathrm{i}\epsilon)} \notag \\
&\phantom{{}={}}\times(2\pi)^4\delta^{(4)}(k-p_1-p_2)\,,
\end{align}
with the dispersion relation~(\ref{eq:dispersion-relation-scalar-mode}) for the unphysical
scalar mode, \eqref{eq:dispersion-relations-longitudinal} for the massive mode, and those
of \eqref{eq:dispersion-relations-transverse} for the transverse modes.
Since the zeroth mode points along the direction of the four-momentum, we are
left with the sum over $\lambda=\pm,3$. We then employ
\begin{align}
&\frac{1}{(k_0-\omega_k^{(\lambda)}+\mathrm{i}\epsilon)(k_0+\omega_k^{(\lambda)}-\mathrm{i}\epsilon)} \notag \\
&\hspace{1cm}=\frac{1}{2\omega_k^{(\lambda)}}\left[\frac{1}{k_0-\omega_k^{(\lambda)}+\mathrm{i}\epsilon}-\frac{1}{k_0+\omega_k^{(\lambda)}-\mathrm{i}\epsilon}\right]\,.
\end{align}
Taking the imaginary part of~\eqref{eq:amplitude-propagator-inserted} by using the general
relation
\begin{equation}
\lim_{\epsilon\mapsto 0^+}\frac{1}{x\pm\mathrm{i}\epsilon}=\mathcal{P}\left(\frac{1}{x}\right)\mp\mathrm{i}\pi\delta(x)\,,
\end{equation}
with the principal value $\mathcal{P}$, results in
\begin{align}
\label{eq:imaginary-part-forward-scattering}
2\text{Im}({\mathcal M}_F)&=e^2\int\frac{\mathrm{d}^4k}{(2\pi)^4}\sum_{\lambda=\pm,3}|J\cdot v^{(\lambda)}|^2\frac{2\pi}{(1+\zeta_1)2\omega_k^{(\lambda)}} \notag \\
&\phantom{{}={}}\times\delta(k_0-\omega_k^{(\lambda)})(2\pi)^4\delta^{(4)}(k-p_1-p_2) \notag \\
&=e^2\int\frac{\mathrm{d}^4k}{(2\pi)^4}\sum_{\lambda=\pm,3}|J\cdot v^{(\lambda)}|^2 \notag \\
&\phantom{{}={}}\times(2\pi)^4\delta^{(4)}(k-p_1-p_2)(2\pi)\delta(\Lambda_{\lambda})\,.
\end{align}
In the final step we exploited that
\begin{align}
\delta(\Lambda_{\lambda})&=\delta\left[(1+\zeta_1)(k_0-\omega_k^{(\lambda)})(k_0+\omega_k^{(\lambda)})\right] \notag \\
&=\frac{1}{(1+\zeta_1)2\omega_k^{(\lambda)}}\left[\delta(k_0-\omega_k^{(\lambda)})+\delta(k_0+\omega_k^{(\lambda)})\right]\,,
\end{align}
for each $\lambda$.
The negative-energy counterparts $k_0=-\omega_k^{(\lambda)}$ do not contribute due to energy-momentum conservation.
The right-hand side of \eqref{eq:imaginary-part-forward-scattering} corresponds to the total
cross section of $\mathrm{e^+e^-}\rightarrow \upgamma$ with both the transverse photon modes and the massive
mode contributing. Also, it is nonnegative under the conditions stated below \eqref{eq:imaginary-part-residues}.
Therefore, we conclude that the optical theorem at tree-level and,
therefore, unitarity are valid for the theory defined by \eqref{eq:lagrange-density-theory}
as long as the photon mass is large enough. Note that the latter computation is generalized to an arbitrary
timelike frame by computing the imaginary part of \eqref{eq:forward-scattering-amplitude-general}
directly with
\begin{equation}
\mathrm{Im}\left(\frac{1}{\Lambda_{\lambda}+\mathrm{i}\epsilon}\right)=-\pi\delta(\Lambda_{\lambda})\,,
\end{equation}
employed. This result means that the decay rate or cross section of a particular process at tree
level can be safely obtained in the context of massive CFJ theory where problems are not expected
to occur (cf.~\cite{Colladay:2016rmy} for the particular example of Cherenkov-like radiation in
\textit{vacuo}).

\section{Application to electroweak sector}
\label{sec:application-electroweak}

Several CFJ-like terms are included in the electroweak sector of the SME before spontaneous
symmetry breaking. We consider the Abelian contribution that is given by \cite{Colladay}
\begin{equation}
\label{eq:cfj-electroweak}
\mathcal{L}_{\mathrm{gauge}}^{\mathrm{CPT-odd}}\supset\mathcal{L}_B\,,\quad \mathcal{L}_B=(k_1)_{\kappa}\varepsilon^{\kappa\lambda\mu\nu}B_{\lambda}B_{\mu\nu}\,,
\end{equation}
with the $\mathit{U}_Y(1)$ gauge field $B_{\mu}$, the associated field strength tensor $B_{\mu\nu}$,
and the controlling coefficients $(k_1)_{\kappa}$. After spontaneous symmetry breaking
$\mathit{SU}_L(2)\otimes\mathit{U}_Y(1)\mapsto \mathit{U}_{\mathrm{em}}(1)$, the field $B_{\mu}$ is
interpreted as a linear combination of the photon field $A_{\mu}$ and the Z boson field $Z_{\mu}$.
The corresponding field strength tensor reads
\begin{equation}
B_{\mu\nu}=F_{\mu\nu}\cos\theta_w-Z_{\mu\nu}\sin\theta_w\,,
\end{equation}
where $\theta_w$ is the Weinberg angle and $Z_{\mu\nu}$ the field strength tensor associated with the
Z boson. Hence, the Lorentz structure of the field operator in $\mathcal{L}_B$ of
\eqref{eq:cfj-electroweak} after spontaneous symmetry breaking has the form
\begin{align}
B_{\lambda}B_{\mu\nu}&=A_{\lambda}F_{\mu\nu}\cos^2\theta_w-(A_{\lambda}Z_{\mu\nu}+Z_{\lambda}F_{\mu\nu})\sin\theta_w\cos\theta_w \notag \\
&\phantom{{}={}}+Z_{\lambda}Z_{\mu\nu}\sin^2\theta_w\,.
\end{align}
Therefore, CFJ-like terms are induced for the massive Z boson as well as for the photon. Our analysis
shows that unitarity issues are prevented in the Z sector due to the mass of this boson that emerges
via the coupling of $Z_{\mu}$ to the vacuum expectation value of the Higgs field. In the photon sector,
a mass has to be added by hand, though.

\section{Conclusions}
\label{sec:conclusions}

In this work, we considered both {\em CPT}-even and {\em CPT}-odd Lorentz-violating modifications for
photons that were constructed from a single preferred spacetime direction. The timelike
sector of the {\em CPT}-odd CFJ electrodynamics is known to exhibit issues with unitarity in the
infrared regime. Therefore, our intention was to find out whether the inclusion of a photon mass can
mitigate these effects.

To perform the analysis, we derived the modified propagator of the theory and decomposed it into a
sum of polarization tensors weighted by the dispersion equations for each photon mode. To get a
preliminary idea on the validity of unitarity, we contracted this propagator with general conserved
currents. The imaginary part of the residue evaluated at the positive poles was found to be nonnegative
as long as the modified dispersion relations for each mode stays real. This property is guaranteed by
the presence of a sufficiently large photon mass. A second more thorough check involved the evaluation
of the optical
theorem for a particular tree-level process. The optical theorem was found to be valid for the same
conditions encountered previously. It is clear how unitarity issues arise in the limit of a vanishing
photon mass when the dispersion relations of certain modes can take imaginary values.

In general, it is challenging to decide whether the imaginary part of the residue of the propagator
contracted with conserved currents is positive for arbitrary preferred directions
and currents. We emphasize that the decomposition of the propagator into polarization tensors allows
for a quite elegant proof of this property independently of particular choices for background fields
and currents. Similar relations are expected to be valuable for showing unitarity in alternative
frameworks.

Hence, we conclude that CFJ electrodynamics is, indeed, unitary when a photon mass is included into
the theory. This finding clearly demonstrates how a mild violation of gauge invariance is capable of
solving certain theoretical issues. Note that unitarity issues are prevented automatically for a
CFJ-like term in the Z-boson sector where the Z boson mass is generated via the Higgs mechanism.

\section*{Acknowledgments}

M.M.F. is grateful to FAPEMA Universal 00880/15, FAPEMA PRONEX 01452/14, and CNPq Produtividade 308933/2015-0.
C.M.R. acknowledges support from Fondecyt Regular project No.~1191553, Chile. M.S. is indebted to FAPEMA
Universal 01149/17, CNPq Universal 421566/2016-7, and CNPq Produtividade 312201/2018-4.
Furthermore, we thank M. Maniatis for suggesting an investigation of the CFJ-like term in the electroweak
sector of the SME.


\end{document}